# Controlling electrical conduction through noble metal thin films by surface plasmon resonance

Suresh C Sharma,[1,*] Vivek Khichar,[1] Hussein Akafzade,[1] Douglas Zinn[1], and Nader Hozhabri[2]

[1]*Department of Physics, University of Texas at Arlington, Arlington, Texas 76019, USA*

[2]*Nanotechnology Research Center, Shimadzu Institute, University of Texas at Arlington, Arlington, Texas 76019*, USA

ABSTRACT

We present evidence in support of a prediction made in the 1950s that the collective oscillations of surface charges (*surface plasmons*) induced by light–matter interactions can influence electrical conduction in two-dimensional noble metals. Two mechanisms were predicted: (1) Umklapp electron–electron scattering and (2) the attractive interaction between free electrons in transition metals because of the screening of the *d*-band electrons by the *s*-band electrons. However, the prediction had not been supported by experiments. We present correlations between the electrical resistivity of noble metal thin films and surface plasmons supporting these theoretical predictions.

[*]Author to whom correspondence should be addressed (sharma@uta.edu)

Since their realization in the 1950s, there has been substantial interest in the physics of collective charge oscillations. The ability to control electrical conduction in thin film devices (for example, in field-effect transistors) by surface plasmon resonance (SPR) offers far-reaching consequences for several different types of application, for example, telecommunications, electronics, and sensors. Bohm and Pines made pioneering contributions to the field in the 1950s [1-4]. Since their early development, plasmon oscillations haves been of immense interest to the condensed matter physics community, particularly *volume plasmons*, and the phenomenon is now well understood.

When energetic electrons pass through a metallic foil, they experience a characteristic energy loss due to the excitation of plasmons in the sea of conduction electrons [5]. The free-electron approximation model of metals, in which high-density free electrons $\left(\sim 10^{23}\,\text{cm}^{-3}\right)$ are treated as a liquid, supports plasma (volume) oscillations propagating through the metal. The frequency of such plasmons is given by

$$\omega_p = \sqrt{\frac{4\pi n e^2}{m_e}}$$

where $n$ is the density of the free electrons, $e$ is the magnitude of the electronic charge, and $m_e$ is the mass of an electron. The plasmon frequency is related to the frequency-dependent dielectric function of the material through

$$\varepsilon(\omega) = \left(1 - \frac{\omega_p^2}{\omega^2}\right)$$

For free-electron metals, the energies of the *volume plasmons* are rather high; they range from approximately 8 eV for Li to about 16 eV for Al [6].

Surface plasmon excitations, on the other hand, are low-energy surface charge density oscillations. The frequency of these high-energy excitations (usually generated by using optical frequencies) is given by

$$\varepsilon(\omega) = \left(1 - \frac{\omega_p^2}{\omega^2}\right)$$

where $n_{2D}$ the two-dimensional density of the occupied surface states [7-11].

Continued interest in surface plasmons has led to the discovery of various fundamental physical processes and applications. For example, (i) the strong electromagnetic fields associated with surface plasmon excitations have contributed to the development of surface-enhanced Raman scattering [12] and (ii) the high sensitivity of SPR to differences in the dielectric properties across thin metal-film/dielectric interfaces has led to the development of numerous sensors [13].

The most commonly utilized nonradiative surface plasmons, excited by optical frequencies, exhibit $\omega \propto \sqrt{q}$ dispersion. On the other hand, low-energy acoustic surface plasmons, which can be excited by long-wavelength infrared sources, have a linear dispersion ($\omega \propto q$). Their energies are just about right to influence the electronic properties of materials. The possibility that these low-energy acoustic plasmons can influence physical processes, like the electrical conductivity of metals, was discussed by Bohm and Pines in the 1950s [1-4]. Pines thought that for the electrical conductivity to be affected, the current had to be changed through electron–electron collisions in two ways: by Umklapp electron–electron scattering or, for highly anisotropic electron energy surfaces, by collisions between electrons of different effective masses. Pines also refers to the possibility of the scattering of *s*-electrons by *d*-electrons in transition metals [4]. The overlap between the partially filled *d*- and *s*-band electrons in transition metals plays an important role in controlling electron conduction in two dimensional noble metals. [14, 15]

Since then, the possibility has been studied by others. For example, Garland considered a two-band model consisting of the *s*-states for the nearly free electrons and *d*-states for the tightly bound electrons [16]. In 1968, Frohlich proposed an alternate mechanism for electron pairing through acoustic plasmons in transition metals [17]. In metals with incomplete *d* bands, two plasmas (*s* and *d*) are present. Under certain conditions, the *d*-plasma is screened by the *s*-electrons, such that, for long wavelengths, its frequencies become proportional to the wave number. It then presents a new acoustic branch, and like the ordinary acoustic branch, this leads to an attractive interaction between the electrons.

Here, we present the results of an experiment designed to probe the influence of surface plasmons on the electrical conductivity (surface resistivity) of noble metal thin films. Our results suggest that there are strong correlations between the collective oscillations of the surface charge density and the electrical conductivity of thin films of noble metals. To what extent they corroborate the theoretical ideas presented in the 1950s and 1960s by Bohm, Pines, Garland, and Frohlich remains to be seen. We believe that our results, the first to our knowledge, will stimulate further experimental and theoretical research on the influence of the collective surface charge density on electron transport processes in materials.

Thin films of high-purity noble metals (purity ≥ 99.999%) were deposited over quartz substrates in a *class-100* clean room for nanofabrication. An AJA International thermal evaporator was used at a base pressure of 1 × 10$^{-6}$ torr with a deposition rate of about 1 Å/s. Three samples were produced: (1) 46-nm-thick gold, (2) 50-nm-thick Ag, and (3) 43-nm-think Cu. Atomic force and scanning electron microscopes were used to examine the surface topography and microstructure of the films. The films were polycrystalline, dense, and smooth at nm scale.

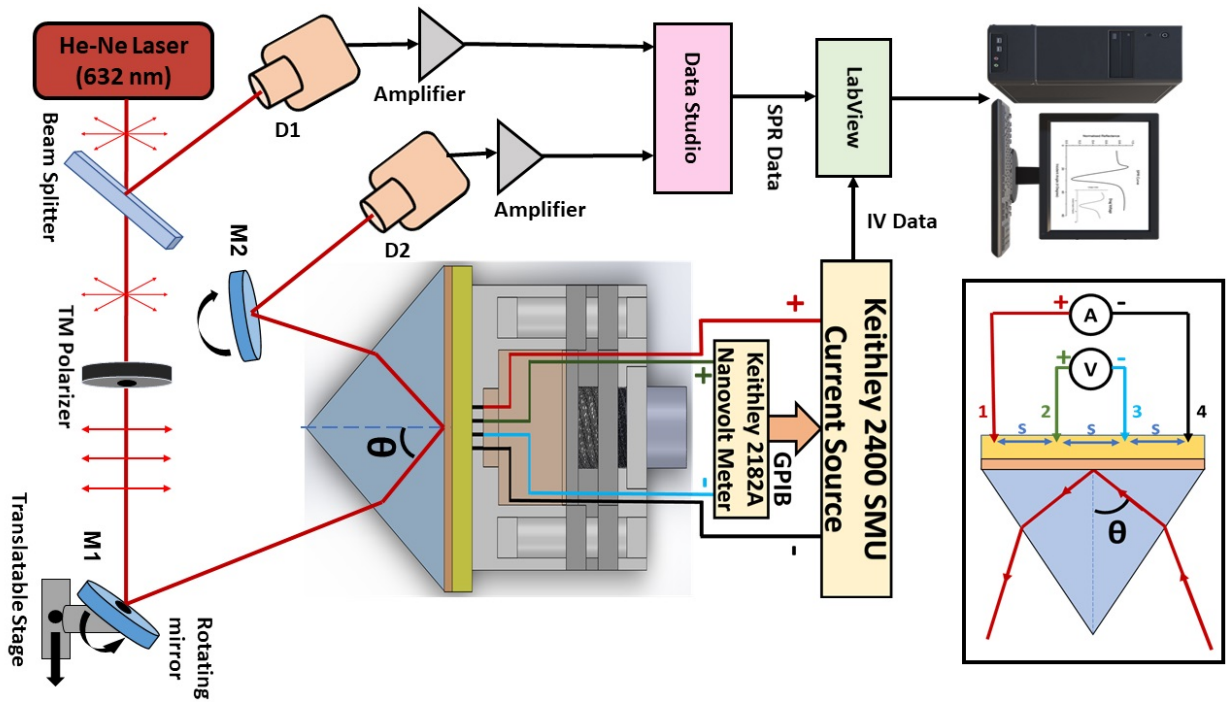

FIG. 1. Block diagram of the experimental system used to measure SPR and electrical resistivity simultaneously.

Figure 1 is a schematic of the system we used to make simultaneous measurements of the SPR and electrical resistivity of the samples. Compared to the traditional Kretschmann configuration optical system [18-21], this system utilizes the recently developed *fixed-detector Kretschmann configuration optical system* for SPR measurements. One of the advantages of the *fixed-detector* system is that it does not require the use of a ($\theta, 2\theta$) goniometer [22]. The angular scans were accomplished by optically steering the incident laser beam. For the electrical resistivity measurements, four contact points on the film were established using a Signatone *four-probe* assembly. This assembly was mounted on a precision micrometer to allow displacement of the contact pins along the normal to the surface of the film. For the four-probe current and voltage measurements, a source meter (Keithley, SMU-2450) was used to apply constant currents to the outer pins of the four-probe assembly and a voltmeter (Keithley, 2182A) was used to record the voltages between the inner two pins of the four-probe unit.

For each angle of incidence, a total of 100 current versus voltage measurements were made for three different current ranges: (1) low current range, from 10 nA to 1 µA, (2) mid current range, 1 µA to 10 µA, and (3) high current range, 10 µA to 100 µA.

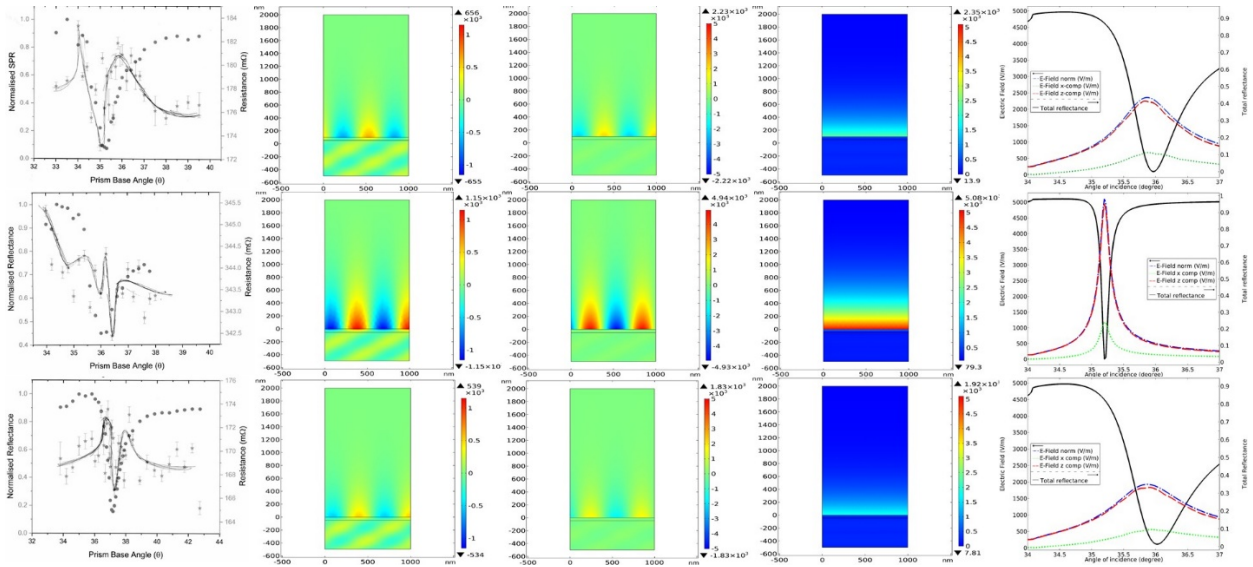

Figure 2. *Results for the Au, Ag, and Cu films: (a) SPR data for* p-polarized laser beam, *(b) electrical resistivity, (c) evanescent field components along the metal/dielectric interface and normal to the interface, and (d) color-coded strengths of the evanescent fields*

Resulting data for the surface electrical resistivity, surface plasmon resonance and COMSOL simulations are shown in figure 2. As far as the surface plasmons are concerned, a characteristic, almost 100% loss in reflectivity, is observed at the resonance angle ($\theta_{SPR} = \sim 35\ nm, for\ Au$). These SPR data are in excellent agreement with computer simulations made with the finite difference method of COMSOL Multiphysics [23, 24]. As seen in figure 2, it is striking that the surface electrical resistivity follows the surface plasmon resonance with the minimum resistivity at the resonance angle. For higher angles, resistivity rises and thereafter asymptotically approaches the resistivity characteristic of the material without surface plasmons. The possibility to control surface resistivity by surface plasmons offers exciting possibilities, e. g., controlling the electron conduction in devices like Field-Effect-Transistors. The evanescent electromagnetic fields of the surface plasmons play an important role; the components in the plane of the film and normal to it are shown in figure 2. It is noteworthy that the fields are extremely high at the surface of the film. For example, the component of the field along and normal to the surface of the film for a 1-mW *p*-polarized incident laser beam used in our experiments is about 2 kV/m. It decays exponentially with distance from the interface with a decay length of approximately 300 nm. The component of the field in the plane of the film is 0.6 kV/m.

In conclusion, we have presented evidence that collective oscillations of the surface charges can influence electrical conduction through the surface layers of noble metal thin films. Our results are consistent with the models proposed by Bohm and Pines, Frohlich, and Garland [1-3, 17, 25-27], whereby Umklapp electron–electron scattering and attractive interactions between free electrons due to screening of the *d*-band electrons by the *s*-band electrons can result in significant changes in the electrical conduction of two-dimensional transition metals. We hope that these results will prompt additional theoretical interest in revisiting the theoretical ideas proposed by Bohm and Pines, Garland, and Frohlich in the 1950-60's.